\documentstyle[preprint,aps,epsfig]{revtex}

\def\be{\begin{equation}}
\def\ee{\end{equation}}

\begin{document}
\draft
\title{Formation of Neutrino Stars from Cosmological Background Neutrinos} 
\author{M. ~H. ~Chan, M.-C. ~Chu}
\address{Department of Physics, The Chinese University of Hong Kong, \\
         Shatin, New Territories, Hong Kong, China }
 
\maketitle

\begin{abstract}
We study the hydrodynamic evolution of cosmological background neutrinos. 
By using a 
spherically symmetric Newtonian hydrodynamic code, we calculate the 
time evolution of the density profiles of 
neutrino matter in cluster and galactic scales. We discuss the 
possible observational 
consequences of such evolution and the resulting density profiles of the 
degenerate neutrino `stars' in galaxies and clusters.

\end{abstract} 
\vskip 5mm
\pacs{PACS Numbers: 95.35.+d, 98.62.Gq, 98.65.Cw}

\section{Introduction}
The observations of neutrino oscillations indicate 
that at least one flavour of neutrinos 
have non-zero rest 
mass, and at least one type of sterile neutrinos with mass of eV order 
must be taken into account for the three 
scales of mass squared differences if the LSND 
experiment is confirmed \cite{Oscillation}. A huge number of primordial 
neutrinos were produced in the early universe \cite{Peebles}, 
\cite{Dodelson}, and they were cooled down  
as the universe expanded, decoupling from baryons much earlier than the 
photon decoupling epoch. 
The temperature of the neutrinos dropped to below 0.1~meV after several 
billions of years, and so if their rest mass $m_{\nu}$ is an eV or so, the 
neutrinos have become non-relativistic. Sterile 
neutrinos may be produced in the early universe by the oscillations of the 
active neutrinos \cite{Dolgov}. These sterile neutrinos may have rest mass 
greater than 10 eV and may play a role in structure formation. Also, 
a neutrino rest mass $m_{\nu}$ of the eV to keV order can account for most 
of the dark matter, which dominates the mass in galaxies and clusters 
\cite{Top}. In this article, we study 
the gravitational collapse of such non-relativisitc cosmic background 
active and sterile neutrinos, and we discuss the formation of degenerate 
neutrino dark halos in cluster and galactic scales. It has been proposed 
that such degenerate neutrino `stars' can be found in 
centers of galaxies \cite{Viollier}, \cite{Apparao}.

\section{Cosmic Background Neutrinos}
In standard cosmology, the neutrinos decoupled at around 1 MeV, and
the present value of their number density $n_{\nu}$ is about 
100 cm$^{-3}$ for each type of neutrinos \cite{Roos}. Therefore, we can 
write the number density of the primordial neutrinos as
\be
n_{\nu} \approx \frac{100}{a^3}~\rm{cm^{-3}},
\ee
where $a$ is the cosmic scale factor, set to be 1 at present. If the 
neutrinos were ultra-relativistic, the neutrino temperature would be
\be
T_{\nu}=\frac{T_0}{a},
\end{equation}
where $T_0 \approx 1.95$ K is a constant \cite{Roos}. When the 
neutrinos cooled down, they became non-relativistic 
and $T_{\nu} \sim a^{-2}$. Suppose at a certain $a=a_c$ and $T_{\nu}=T_c$, 
$kT_c=m_{\nu}c^2$, and 
the neutrinos become non-relativistic. We have 
\be
T_c \approx \frac{T_0}{a_c} \approx \frac{T_1}{a_c^2},
\end{equation}
where $T_1$ is the present neutrino temperature. Therefore, we can express
$T_{\nu}$ in terms of $m_{\nu}$(in eV):
\be
T_{\nu} \approx \frac{3.3 \times 10^{-4}}{m_{\nu}a^2} \rm{K}.
\end{equation}
for the non-relativistic 
regime. We will use this relation to check the degeneracy of the neutrinos 
at the structure formation epoch in the next section.

In order to estimate the time for the beginning of structure formation, we 
should consider the Jeans mass of the primordial neutrinos, which is 
defined as
\be
M_J= \frac{4 \pi}{3} \rho_{\nu} \lambda_J^3,
\ee
where $\rho_{\nu}$ is the mass density of the neutrinos and $\lambda_J$ is 
called the Jeans length which is defined by
\be
\lambda_J=\left(\frac{\pi c^2}{G \bar{\epsilon}} \frac{ \partial 
P_{\nu}}{\partial \rho_{\nu}} \right)^{1/2},
\ee
with $\bar{\epsilon}$ the mean energy density and $P_{\nu}$ the 
pressure of the neutrinos. When the mass of an object is larger than its 
Jeans mass, the object will collapse gravitationally to form structure. 
The mass density of the neutrinos in the zero 
$T_{\nu}$ limit is given by
\be
\rho_{\nu}=\frac{K}{c^2} \left[x_F(2x_F^2+1)\sqrt{1+x_F^2}- \ln \left(x_F+ 
\sqrt{1+x_F^2} \right) \right],
\ee
where $K$ is a constant and $x_F$ is defined as 
\be
x_F= \frac{2 \pi \hbar}{m_{\nu}c^2} \left[\frac{3n_{\nu}}{8 \pi} 
\right]^{1/3}.
\ee
$x_F \ll 1$ and $x_F \gg 1$ represent the non-relativistic and 
ultra-relativistic regimes respectively for the neutrinos.
By combining Eqs.~(5)-(8), we obtain a relation between 
$M_J$ and $a$ for a given $m_{\nu}$ (see Fig.~1). If the Jeans mass is of 
galactic scale (about $10^{12}M_{\odot}$), the value of $a$ is about 
0.0585 for $m_{\nu}=10$ eV but greater than one for $m_{\nu}=1$ eV. For
cluster scale (about $10^{15}M_{\odot}$), the corresponding values of $a$ 
are $6 \times 10^{-4}$ and 0.0125 for $m_{\nu}$ being 10 eV and 1 eV 
respectively. 

By looking at the Jeans mass only, we find that clusters 
will start to form first and then the galaxies, which is known as 
the top-down scenario. This disagrees with 
the observation, which indicates that the formation of smaller 
structures occurred earlier in time \cite{Top}. However, we should 
compare not only the time that the structures start to form, but also the 
duration of the formation of structures in order to determine 
whether neutrinos can still form structures and agree with the 
observation. 

We believe that different 
masses of neutrinos may correspond to different structures- the larger the 
neutrino mass is the smaller the structure and vice versa. The radius of a
hydrostatic neutrino star is given by \cite{Radius}
\be
R=20.7~{\rm{pc}} \left(\frac{M}{10^6M_{\odot}} \right)^{-1/3} 
\left(\frac{m_{\nu}}{1~\rm{keV}} \right)^{-8/3},
\ee
where $M$ is the total mass of the neutrino star. The corresponding $R$
are Mpc, kpc and 0.01 pc for $m_{\nu}=1$ eV, 10 eV and 
1 keV respectively if $M=10^{15}M_{\odot}$. Therefore, we assume that 
neutrinos with rest mass of 10 eV or above will 
dominate in galaxy scale and those with rest mass
1 eV or below will form structures in cluster scale.
  
\section{Hydrodynamics of Neutrino Star Formation}
The formation of degenerate heavy 
neutrino stars was discussed in Ref.~10 using the analogy with an 
interacting self-gravitating Bose condensate. Here, 
we use a hydrodynamical code to simulate the formation 
of neutrino stars in galaxies and clusters. We use the 
Lagrangian mass coordinate to write the spherically symmetric 
hydrodynamic equations \cite{Bowers}: 
\be
 \frac{ \partial u}{ \partial t}=-4 \pi r^2 \left[ \frac{ \partial P}{ 
\partial m}+ \frac{ \partial Q}{ \partial m} \right]- \frac{Gm(r)}{r^2},
\ee
\be
 \frac{1}{ \rho}=4 \pi r^2 \frac{ \partial r}{ \partial m}.
\ee
Here, $u$ is the fluid velocity, $ \rho$ the mass density, $m$  
the enclosed mass, and $Q$ is the artificial viscous stress. The 
above equations indicate that the neutrinos are under gravitational 
attraction and influence of the pressure $P$. There exists neutrino 
degeneracy pressure as the neutrinos become 
degenerate after they have gravitationally collapsed. The criterion of
degeneracy is given by
\be
\rho_{\rm{deg}} \geq m_{\nu} \left(\frac{m_{\nu}kT_{\nu}}{2 \pi \hbar^2} 
\right)^{3/2}.
\ee
We can use Eq.~(4) to approximate $T_{\nu}$ and set a lower limit for the 
mass density in degenerate state.
The neutrino degeneracy pressure is given by
\be
P_{ \nu}= \frac{4 \pi^2 \hbar^2}{5m_{ \nu}} \left( \frac{3}{4 \pi} 
\right)^{2/3}n_{ 
\nu}^{5/3}. 
\ee
In fact, the neutrinos are collisionless 
except for the degeneracy pressure. Therefore, we set the effective 
temperature of the neutrinos to be 
zero because the neutrino temperature is much less than their Fermi energy. 
Here, we consider non-relativistic degenerate neutrinos, the Fermi 
momentum for which is much smaller than their rest mass.   
In the hydrodynamic code, we use an artificial viscous stress to 
smooth out possible shock waves. The artificial viscous stress is defined as:
\be
Q=- \alpha \left[ \frac{ \partial u}{ \partial r}- \frac{1}{3r^2} \frac{ 
\partial}{ \partial r}(r^2u) \right],
\ee
where $ \alpha$ is the scalar artificial viscosity defined by 
\be
 \alpha=l^2 \vert { \frac{ \partial \rho}{ \partial t}} \vert,
\ee
with $l$ being a constant length which will be defined later.
In the hydrodynamic simulation, we use a finite difference scheme to 
evaluate the above equations \cite{mphil}. We represent all the variables as:
$u \rightarrow u_k^{n+1/2}, r \rightarrow r_k^{n+1/2}, \rho \rightarrow 
\rho_{k+1/2}^n, P \rightarrow P_{k+1/2}^n$, where the upper index 
represents time step and the lower index labels mass shell, and we let 
\be
l^2=C_Q( \Delta r_{k+1/2})^2,
\ee
where $ \Delta r_{k+1/2}$ is the thickness of the $ \rm{k}^{ 
\rm{th}}$ mass shell and $C_Q$ is an adjustable parameter set to be 
between 2 to 3, to 
spread the shock front into several zones. The difference between two 
consecutive Lagrangian coordinates 
is the mass of a zone, $ \Delta m \rightarrow \Delta m_{k+1/2}=m_{k+1}-m_k$.
In addition we use the Courant stability condition and test for 
convergence to make sure that the computer code works effectively.

In our simulations, we set the initial density profiles to be uniform. We 
use $10^{12}M_{\odot}$ and $10^{15}M_{\odot}$ as the Jeans 
mass to find the values of $a$ at which gravitational collapse begins for 
different $m_{\nu}$. The initial density $\rho_0$ can be obtained from 
the standard cosmology:
\be
\rho_0=\frac{ \Omega_{\nu} \rho_c}{a^3},
\ee
where $\Omega_{\nu}$ and $\rho_c$ are the cosmological density parameter 
of the 
neutrinos and critical density of the universe respectively. All the 
parameters used in the simulations are shown in Table 1. 
Also, for simplicity, the initial hydrodynamic velocity of the 
neutrinos is set to be zero.

In fact, after the epoch of decoupling, the neutrinos were 
not fully degenerate and had no interaction with other neutrinos, baryons, 
or photons. Therefore, the pressure is very small and not 
enough to prevent gravitational collapse. However, when the neutrinos 
continue to 
collapse, their density will increase until they become fully 
degenerate. Further increase of the 
number density will boost up their degeneracy pressure until the pressure 
gradient balances the 
gravitational attraction and the collapse is stopped. 
If the neutrino mass is too large, the degeneracy pressure will be small and 
the neutrinos will collapse to a high density within a very small 
region with a high speed. Therefore, the neutrinos will rebound back with 
a high velocity. If the neutrino mass is too small, the 
neutrinos would remain relativistic and the gravitational 
attraction is 
not enough to bind them together. In other words, there only 
exists a range of neutrino mass to form a stable neutrino star. 
In Fig.~2-5, we demonstrate three cases with the parameters in Table 1. 
In Fig.~2, the resulting density becomes smaller as time evolves, 
which 
means that the neutrinos cannot form structures by themselves if $m_{\nu}$ 
is 1 eV. In Fig.~3, we still use $m_{\nu}=1$ eV but we impose a constant 
density core ($10^{-24}$ gcm$^{-3}$ within 500 kpc) inside the cluster. 
The neutrinos can form structure with the
help of additional mass.  In Fig.~4 and 5, we use $m_{\nu}=10$ eV and 1 
keV, and the neutrinos can form structures by themselves. 
The total mass of the structures we used are $10^{15}M_{\odot}$ for 
$m_{\nu}=1$ eV and $10^{12}M_{\odot}$ for $m_{\nu}$=10 eV or 1 keV. 
These few figures show the possibility of neutrinos forming structures in 
both galaxies and clusters. 
Also, the time scale of formation is smaller as $m_{\nu}$ is 
larger. The time scales are about $10^{12}$ s, $10^{16}$ s and $10^{17}$ s 
if $m_{\nu}$ are 1 keV, 10 eV and 1 eV respectively. Therefore, 
when one includes 
the formation time, the sequence of neutrino structure formation 
agrees with the observation, i.e. smaller scales form first.

In Fig.~6, we compare the density profile at hydrostatic 
equilibrium with the late time profile of the dynamical evolution. They 
appear almost identical except at the outermost region, where the 
difference arises because of 
some undamped oscillations in the outer part of the neutrino star. 
The density profile is nearly 
constant at small radius and drops greatly at large radius. 

In cluster formation, most of the baryons become hot gas particles and 
stars in galaxies. Their mass is about one-tenth of the total mass of the 
cluster. We have ignored the effect of baryons here in the simulation. 
Nevertheless, the constant density profile at small radius agrees with 
some observational data of cluster dark matter \cite{Tyson}, 
\cite{Manho}.

When the neutrinos begin to collapse by gravitational attraction, they 
gain kinetic energy 
and the density of the inner part of the neutrino star increases. When 
the central density 
is high enough, the pressure gradient prevents the neutrinos to collapse 
further and some of the neutrinos begin to rebound back. 
We find that there are oscillations in the neutrino star. 
In Fig.~3, we can see oscillations of the central density. Also in 
Fig.~7, we plot the density against time at $r=50$ 
kpc with the same parameters as those used in Fig.~4. The period of 
oscillation $ \tau$ is almost constant and the 
amplitude varies with time. The value $ \tau$ in the 
central region can be found by Taylor expansion of the hydrodynamic 
equation with small displacement, which is roughly given by the following 
relation \cite{manhomphil}: 
\be
 \tau \approx \frac{2 \pi}{ \sqrt{4 \pi G \rho_f/3}}.
\ee
From Fig.~7, the period of oscillation is about 1.6 billion years. The 
density at $r=50$ kpc is about $6 \times 10^{-26}$~gcm$^{-3}$. Therefore, 
by Eq.~(18), $ \tau \approx $ 1.5 billion years, which is roughly 
consistent 
with the result. Because of the weak coupling among the neutrinos, the 
oscillations probably would not be damped out for a long time.

\section{Discussion}
We have demonstrated the possibility that the neutrinos 
can form structures in galaxies and clusters. Including the 
formation time, our model is consistent with the observation but 
not the top-down scenario. Our hydrodynamical simulation results 
(see Fig.~4) agree with recent observed data 
indicating that the central density of dark matter profile is 
constant \cite{Tyson}. 

Light neutrinos (1 eV) cannot form structure by their own gravitational 
attraction. However they 
can still form structures under the gravitational field of a heavier 
component (see Fig.~2 and 3). 
It is well known that light primordial neutrinos (eV order) cannot account 
for all the dark matter as $\Omega_{\nu}$ is much less than 0.3 for 
$m_{\nu}=1$ eV \cite{Hu}. There must exist heavier particles such as 
sterile neutrinos which 
dominate the mass of the dark matter and provide such a gravitational 
field 
in the clusters. These sterile neutrinos (10 eV order) may also form 
smaller structures in galaxies (about 10-20 kpc), and all these smaller 
structures link up 
together to form a large structure in clusters. Therefore, neutrinos 
with rest mass 10 eV or above are important in structure formation, and 
we should not neglect their contribution. The range of possible 
masses of active neutrinos and sterile neutrinos can both be constrained 
by more data from the distribution of hot gas in clusters and the rotation 
curves of galaxies.

From Fig.~3 and 7, we can see that the central density 
of a neutrino star oscillates, 
which may affect the hot gas in 
clusters. It may force oscillate the hot gas, which may be observable. 
Also, the oscillations in the 
hot gas may convert the kinetic energy into internal energy of the hot gas 
and thus heat them up, which may at least partially 
account for the heat source of the hot gas. 

\section*{Acknowledgments}
The work described in this paper was substantially supported by a 
grant from the Research Grants Council of the Hong Kong Special 
Adminstrative Region, China (Project No. 400803).

\begin{table}[t]
\caption{The parameters used in the simulations.}
\label{table}
\end{table}
\begin{center}
\begin{tabular}{|c|c|c|}
\hline
$m_{\nu}$&$\rho_0$(gcm$^{-3}$)&Cut off radius (kpc) \\
\hline
1 eV&$1.02 \times 10^{-25}$&539 \\
\hline
10 eV&$9.97 \times 10^{-27}$&11.7 \\
\hline
1 keV&$1.41 \times 10^{-18}$&0.022 \\
\hline
\end{tabular}
\end{center}

\begin{figure}[h]
\psfig{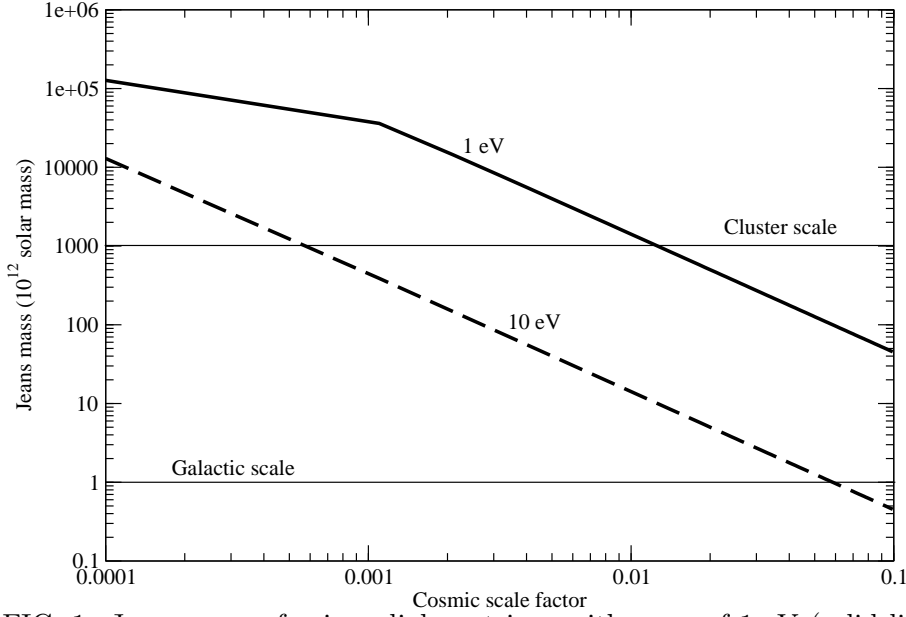}
\caption{Jeans mass of primordial neutrinos with mass of 1 eV (solid 
line) and 10 eV (dashed line) respectively vs. cosmic scale factor. Also, 
indicated are typical cluster and galactic mass scales.}
\end{figure}
\vskip 10mm

\begin{figure}
\psfig{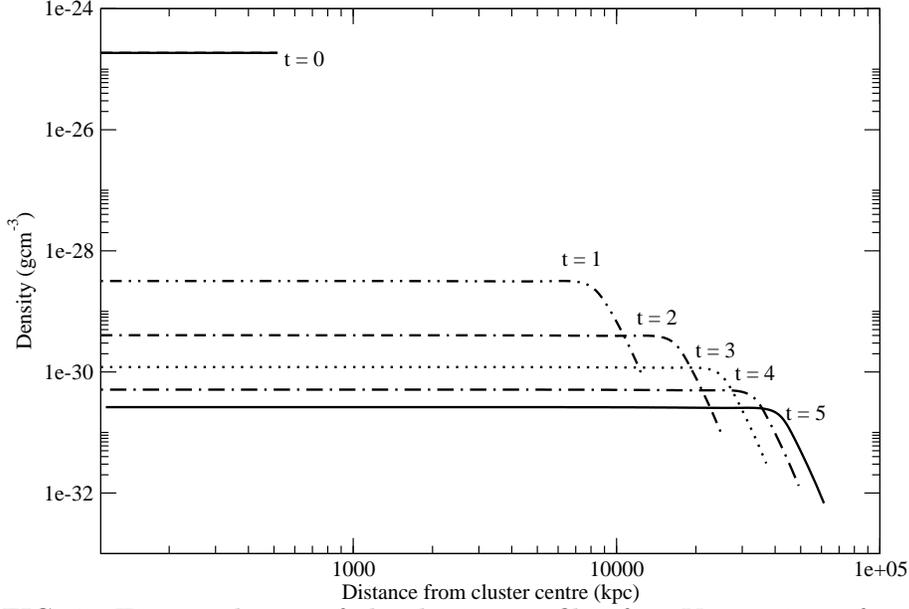}
\caption{Time evolution of the density profile of 1 eV neutrinos, from 
initially uniform and stationary distribution. Total 
mass = $10^{15}M_{\odot}$ and the time steps are in $10^9$ years.}
\end{figure}
\vskip 10mm

\begin{figure}
\psfig{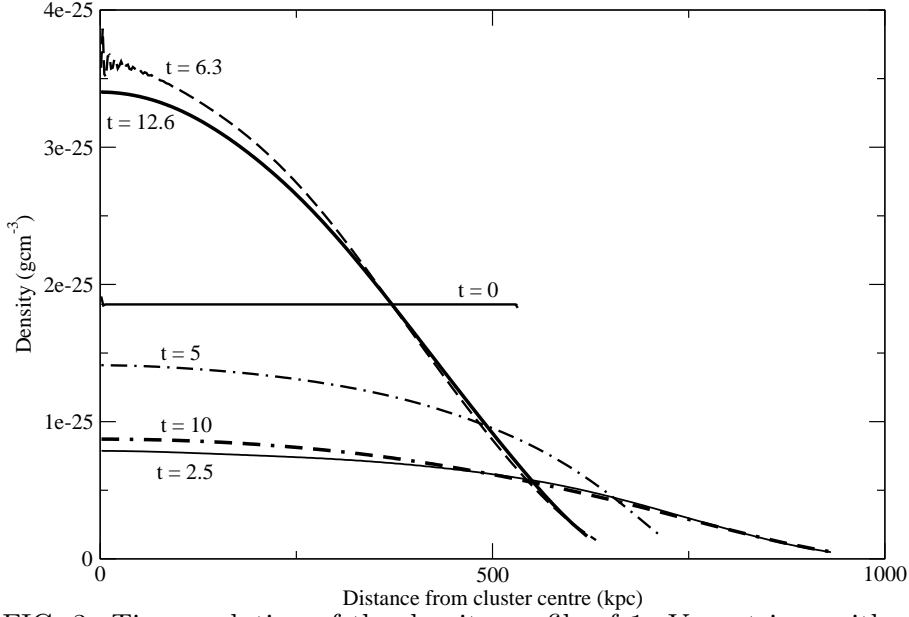}
\caption{Time evolution of the density profile of 1 eV neutrinos with 
a constant density core ($\sim 10^{-24}$ gcm$^{-3}$ within 500 kpc) of 
dark matter and 
total mass $10^{15}M_{\odot}$ in the background. The time steps are in 
$10^9$ years.}
\end{figure}
\vskip 10mm

\begin{figure}
\psfig{file=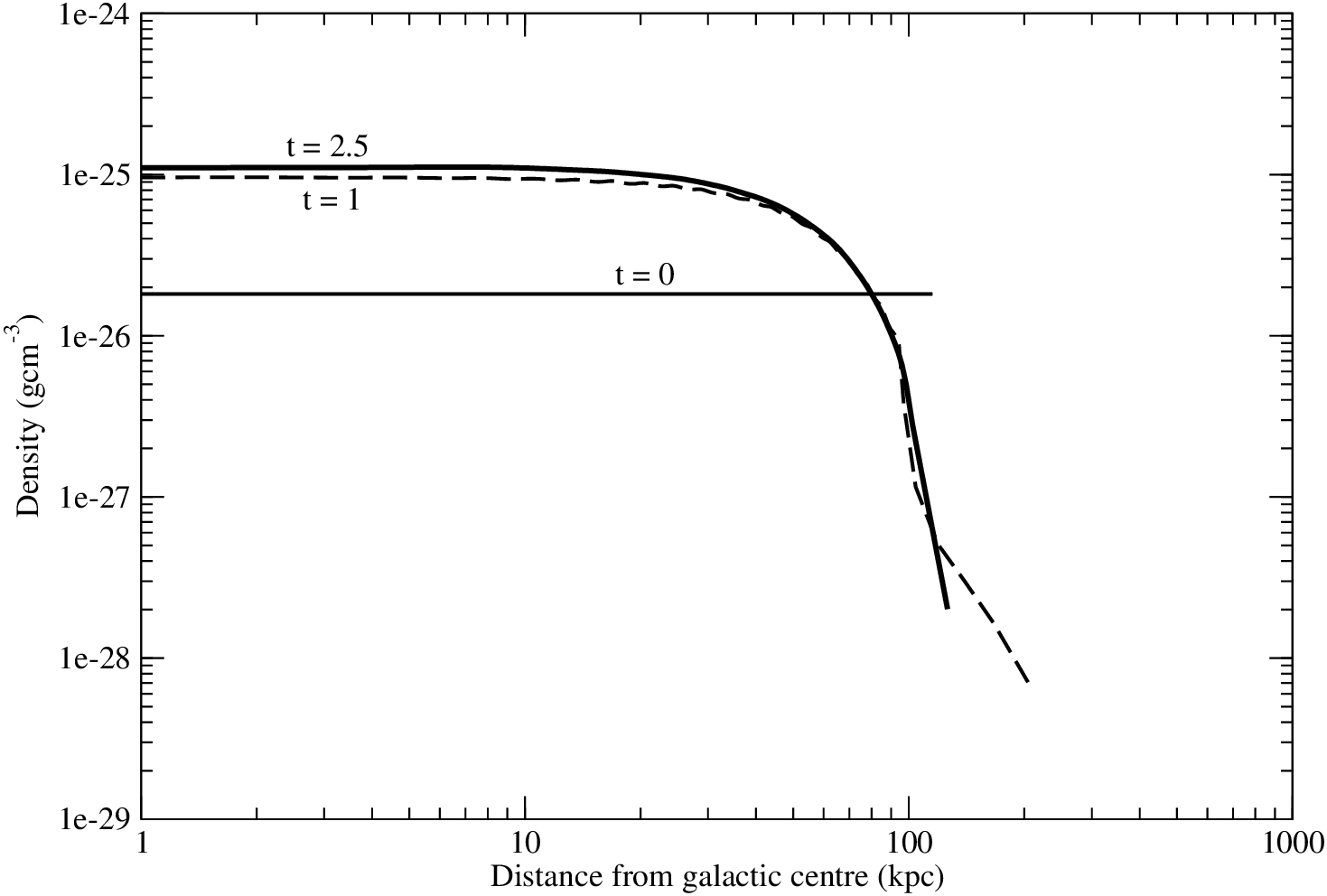,angle=0,width=12cm}
\caption{Time evolution of the density profile of 10 eV neutrinos. Total 
mass = $10^{12}M_{\odot}$. The time steps are in $10^9$ years.}
\end{figure}
\vskip 10mm

\begin{figure}
\psfig{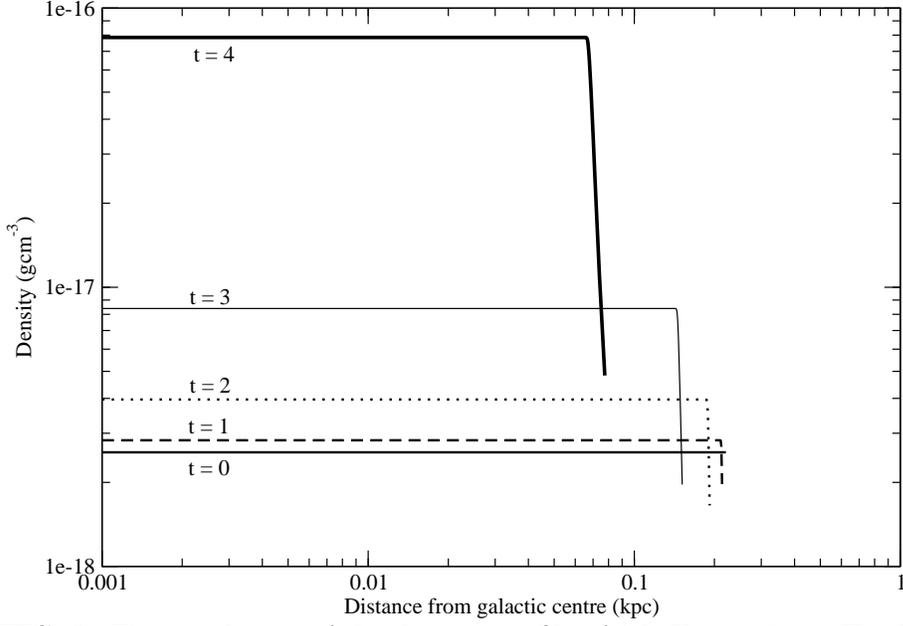}
\caption{Time evolution of the density profile of 1 keV neutrinos. Total 
mass 
= $10^{12}M_{\odot}$. The time steps are in $10^4$ years.} 
\end{figure}
\vskip 10mm

\begin{figure}
\psfig{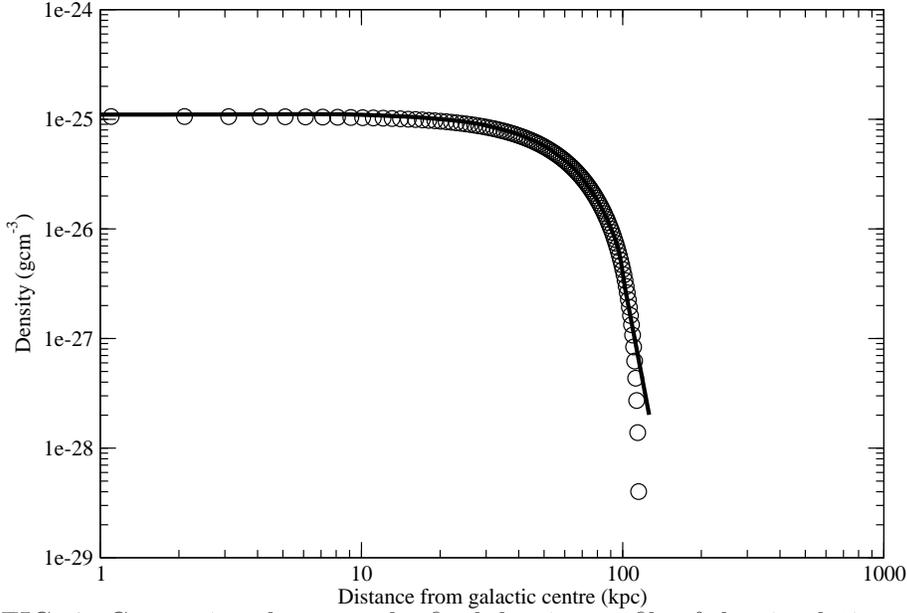}
\caption{Comparison between the final density profile of the simulation 
and hydrostatic density profile with the same parameters as those used in 
Fig.~4. The solid line represents the final 
density profile of the hydrodynamic simulation. The circles represent the 
density profile of the hydrostatic neutrino star.} 
\end{figure}
\vskip 10mm

\begin{figure}
\psfig{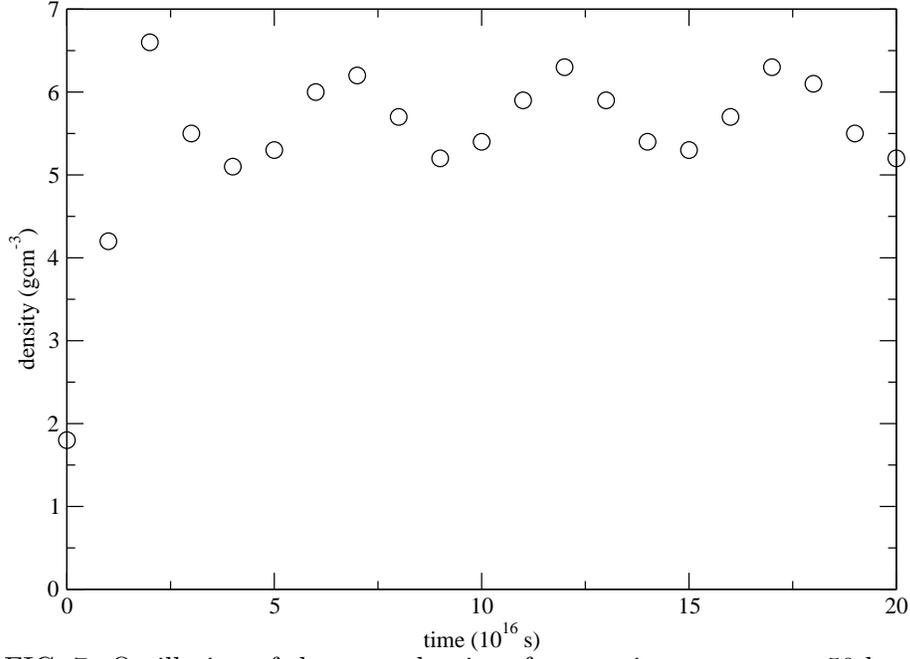}
\caption{Oscillation of the mass density of a neutrino star at $r=50$ kpc 
with the same parameters as those used in Fig.~4.} 
\end{figure}
\vskip 10mm 


\begin{references}
\bibitem{Oscillation} S.~M.~Bilenky, C.~Giunti and W.~Grimus, 
Eur.~Phys.~J.~C {\bf 1}, 247 (1998).

\bibitem{Peebles} See for example, P.~J.~E.~Peebles, {\it Principles of 
Physical Cosmology} (Princeton University Press, Princeton, 1993).

\bibitem{Dodelson} S.~Dodelson and L.~M.~Widrow, Phys.~Rev.~Lett., 
{\bf 72}, 17 (1994).

\bibitem{Dolgov} A.~D.~Dolgov and S.~H.~Hansen, hep-ph/0009083 v3 (2001).

\bibitem{Top} B.~S.~Ryden, {\it Introduction to Cosmology} 
(Addison-Wesley, 2003).

\bibitem{Viollier} R. ~D. ~Viollier, N. ~Bilic and F. ~Munyaneza, Phys. 
Rev. D {\bf 59},1 (1998).

\bibitem{Apparao} K.~M.~V.~Apparao, astro-ph/0204375 v1 (2002).

\bibitem{Roos} M.~Roos, {\it Introduction to Cosmology} (John Wiley and 
Sons, 1994).

\bibitem{Radius} R.~Bowers and T.~Deeming, {\it Astrophysics I Stars} 
(Jones and Bartlett, 1984).

\bibitem{Bilic} N.~Bilic, R.~J.~Lindebaum, G.~B.~Tupper and 
R.~D.~Viollier, astro-ph/0106209 v2 (2001).

\bibitem{Bowers} R.~L.~Bowers and J.~R. ~Wilson, {\it Numerical 
Modelling in Applied Physics and Astrophysics}, Chapter 4 (Jones and
Bartlett, 1991). 

\bibitem{mphil} T.~W.~Lee, {\it Computational Study of Type II Supernova 
Explosion} (M.Phil. Thesis, CUHK, 1999).

\bibitem{Tyson} J.~A.~Tyson, G.~P.~Kochanski and I.~P.~Dell'Antonio, ApJ. 
{\bf 498}, L107 (1998). 

\bibitem{Manho} M.~H.~Chan and M.-C.~Chu, astro-ph/0401329 (2004).

\bibitem{manhomphil} M.~H.~Chan, {\it Properties and Formation of 
Neutrino Star} (M.Phil. Thesis, CUHK, 2004).

\bibitem{Hu} W.~Hu, D.~J.~Eisenstein and M.~Tegmark, Phys.~Rev.~Lett.~ 
{\bf 80}, 5255 (1998).
\end{references}
\end{document}